\documentclass[10pt]{IEEEtran}
\IEEEoverridecommandlockouts
\usepackage{cite}
\usepackage{amsmath,amssymb,amsfonts}
\usepackage{algorithmic}
\usepackage{graphicx}
\usepackage{textcomp}
\usepackage{xcolor}
\usepackage{balance}
\usepackage{hyperref}
\usepackage[nocomma]{optidef}
\usepackage[scientific-notation=true]{siunitx}
\usepackage{adjustbox}
\usepackage{subcaption}

\def\BibTeX{{\rm B\kern-.05em{\sc i\kern-.025em b}\kern-.08em
    T\kern-.1667em\lower.7ex\hbox{E}\kern-.125emX}}
    
\begin{document}

\title{Goal-oriented Communications for the IoT: \\System Design and Adaptive Resource Optimization\vspace{.2cm}}

\author{Paolo Di Lorenzo, Mattia Merluzzi, Francesco Binucci, Claudio Battiloro, Paolo Banelli,\smallskip\\ Emilio Calvanese Strinati, and Sergio Barbarossa  

\thanks{   

This work was supported by the European Union under the Italian National
Recovery and Resilience Plan (NRRP) of NextGenerationEU, partnership on
“Telecommunications of the Future” (PE00000001 - program “RESTART”), by the ANR under the France 2030 program, grant "NF-NAI: ANR-22-PEFT-0003", and by MIUR under the PRIN 2017 Liquid Edge project.  Di Lorenzo, Battiloro, and  Barbarossa are with Sapienza University of Rome, Italy (emails: \{paolo.dilorenzo, claudio.battiloro, sergio.barbarossa\}@uniroma1.it). Mattia Merluzzi and Emilio Calvanese Strinati are with the Univ. Grenoble Alpes, CEA, Leti, Grenoble, France (e-mail:
\{mattia.merluzzi, emilio.calvanese-strinati\}@cea.fr).
Binucci and  Banelli are with the University of Perugia,  Italy (e-mail:
\{francesco.binucci, paolo.banelli\}@unipg.it).} \vspace{-.4cm}}

\maketitle


\begin{abstract}
Internet of Things (IoT) applications combine sensing, wireless communication, intelligence, and actuation, enabling the interaction among heterogeneous devices that collect and process considerable amounts of data. However, the effectiveness of IoT applications needs to face the limitation of available resources, including spectrum, energy, computing, learning and inference capabilities. This paper challenges the prevailing approach to IoT communication, which prioritizes the usage of resources in order to guarantee perfect recovery, at the bit level, of the data transmitted by the sensors to the central unit. We propose a novel approach, called goal-oriented (GO) IoT system design, that transcends traditional bit-related metrics and focuses directly on the fulfillment of the goal motivating the exchange of data. The improvement is then achieved through a comprehensive system optimization, integrating sensing, communication, computation, learning, and control. We provide numerical results demonstrating the practical applications of our methodology in compelling use cases such as edge inference, cooperative sensing, and federated learning. These examples highlight the effectiveness and real-world implications of our proposed approach, with the potential to revolutionize IoT systems.

\end{abstract}

\section{Introduction and Related Works}

IoT applications are fueling artificial intelligence (AI) engines with increasingly complex and high dimensional data collected thanks to sensing, actuation, processing, and wireless communication capabilities of a plethora of heterogeneous physical devices. 
IoT systems are leading to a seamless integration between the interconnected physical world of sensors, devices and actions, and its programmable digital representation, forming a cyber-physical continuum with advanced intelligence and unrestricted connectivity. To enable this vision, IoT networks must rely on an efficient connect-compute infrastructure that is \textit{AI-native}, i.e., inherently built on AI tools \cite{letaief2021edge}. 
On the one hand, AI algorithms will be leveraged for network optimization and orchestration; on the other hand, AI will be made available as a service at the network edge (a.k.a., Edge AI) thanks to a robust and efficient communication platform capable of processing and extracting valuable information from vast amounts of raw data as close as possible to where the data are produced and used. However, IoT networks impose stringent and unique requirements in terms of effectiveness (e.g., latency, AI performance), since communications often require the exchange of massive and high-dimensional data between devices having markedly different natures (e.g., humans and machines).
As an example, considering time-critical IoT applications such as, e.g., autonomous driving, industrial automation and control, and smart surveillance, it is indispensable not only to provide (distributed) AI services with high reliability (e.g., high dependability and trustworthiness), but also within ultra-low latency, while taking into account various factors such as energy efficiency, communication overhead, computation capabilities, and more.  
Despite ongoing efforts, the current trend in IoT systems is characterized by a growing demand for wider bandwidth, as well as increased energy consumption, to accommodate the ever-rising need of higher data rates, driven by emerging services like virtual reality and autonomous driving. 
The relentless pursuit of these requirements necessarily faces a bottleneck caused by resource scarcity, including limited spectrum, energy availability, and computing power, thus requiring the development of a new communication paradigm to support future IoT applications.


A fundamental paradigm shift for envisioned IoT applications is given by the \textit{semantic and goal-oriented communications} approach \cite{strinati20216g}, which goes beyond the typical bit-related metrics that are used today in system design and optimization, focusing instead on the recovery of the meaning conveyed by the transmitted bits and/or the effective fulfillment of the tasks motivating the exchange of information. 
For instance, GO data compression is a key element of this new design \cite{zhang2022goal,stavrou2023role}, whose aim is to extract only relevant and useful information to the end-users/applications for serving the decision-making at the receiver with the required accuracy and respecting strict time constraints.
Very recently, there was a surge of interest in GO and semantic communications  \cite{strinati20216g, gunduz2022beyond,stavrou2023role,zhang2022goal,kountouris2021semantics,xie2021deep}. 
Several aspects need to be addressed to achieve efficient GO communications, including model compression techniques, adaptive model selection/splitting strategies, and application-specific optimizations, 
see, e.g., \cite{letaief2021edge}. In \cite{kountouris2021semantics}, the authors propose semantics-empowered sampling and communication policies to a communication scenario in which the destination is tasked with real-time source reconstruction for the purpose of remote actuation. Goal-oriented adaptive quantization and compression schemes were also proposed in \cite{MerluzziEML2021}, \cite{shlezinger2021deep} to explore the rate-performance trade-off. A principled data compression strategy matched to the learning task and based on the (variational) information bottleneck has been also proposed in \cite{shao2021learning}.
Finally, a popular approach for semantic and GO communications is joint source channel coding \cite{jankowski2020wireless}, which was proved to outperform traditional separate source and channel coding in the case of limited delay and complexity.

\textbf{Contributions.} The aim of this paper is to put forward a GO system design for the IoT, encompassing the key aspects related to architecture design and GO adaptive resource optimization. 
Differently from several other papers on GO and semantic communications, which
mainly focus on the design of proper data compression/coding schemes, e.g., \cite{zhang2022goal,stavrou2023role,gunduz2022beyond,xie2021deep,shlezinger2021deep,shao2021learning,jankowski2020wireless}, our paper looks at the problem from a wider and cross-layer perspective.
Starting from our definition of goal effectiveness and the proposed IoT architecture, we formally define a communication goal as the fulfillment of a task (e.g., learning, actuation, control, etc.) with a target effectiveness level, which can be typically expressed in terms of a set of cross-layer performance measures (e.g., task accuracy, E2E latency). This approach naturally leads to a joint system optimization encompassing sensing, communication, computation, learning and control aspects, with the final aim of achieving effective GO communications with a minimum cost, e.g., in terms of energy or resource consumption. Such formulation entails striking the desired trade-off between goal effectiveness and cost, which will be dynamically explored adapting the available degrees of freedom to cope with several sources of randomness affecting the IoT system such as, e.g., wireless channels, data arrivals, data distribution over space and time, CPU availability, energy harvesting, etc. We will leverage the interplay between model-based optimization (e.g., network stochastic optimization), which enable continuous learning and adaptation in random environments, and purely data-driven approaches (e.g., deep reinforcement learning) that are useful either when the environment is totally unknown, or it is too complex to be modeled. 
Finally, we apply the proposed GO system design and adaptive optimization to the following three use-cases: \textit{i)} \textit{GO compression for edge inference}, where several IoT devices transmit compressed features to perform real-time classification with latency and accuracy constraints; \textit{ii)} \textit{Cooperative effective sensing}, where the goal of the sensor network is to reconstruct a signal field, under mean-square error constraints, sending compressed and noisy data to a fusion center (FC); \textit{iii)} \textit{GO federated learning}, where a set of devices aims to train a common deep neural network architecture with E2E delay and accuracy constraints, exchanging compressed models computed from locally collected data.

\section{Goal-oriented IoT Architecture Design}
\begin{figure*}
    \centering
    \includegraphics[width=.98\textwidth]{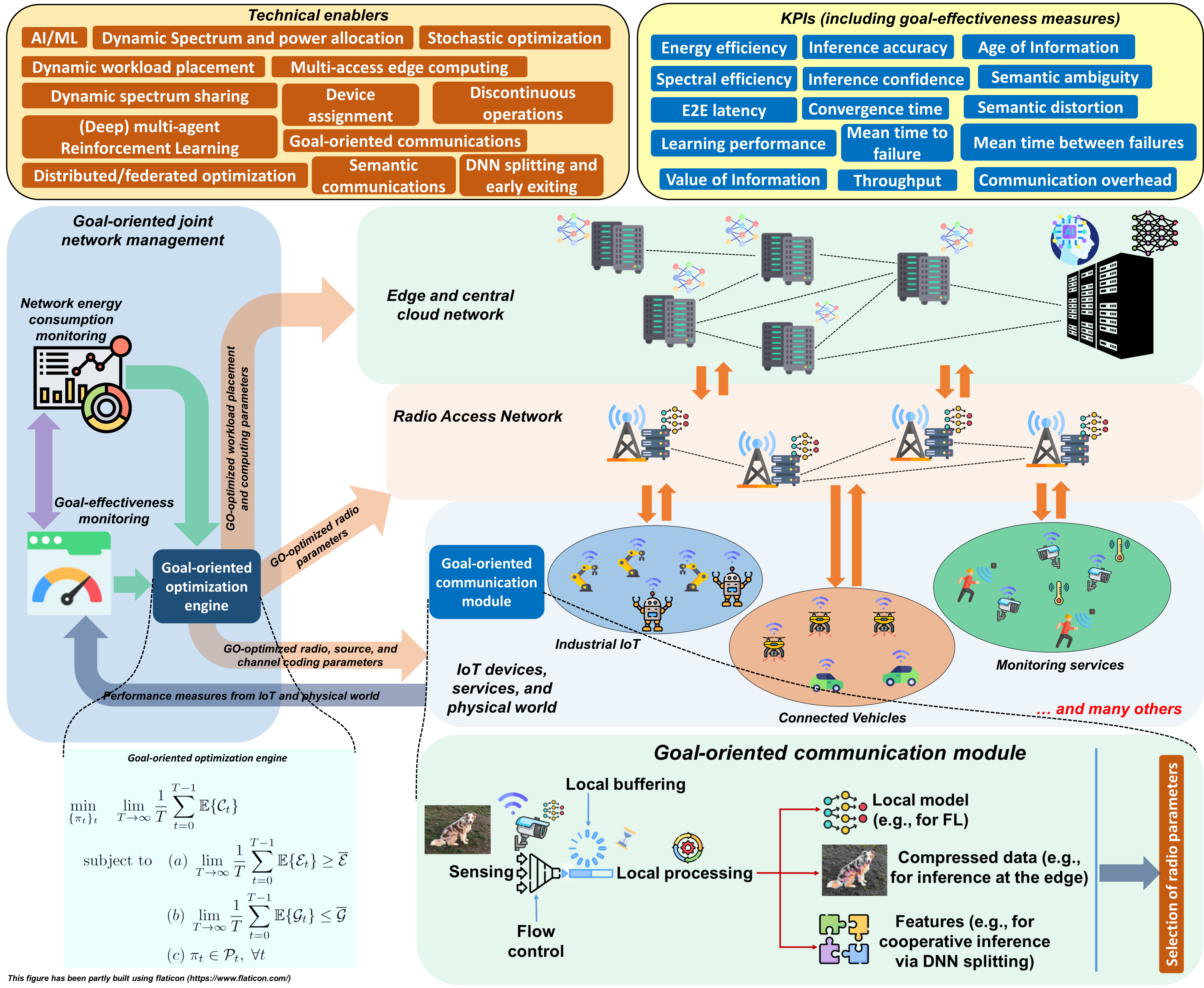}
    \caption{Architecture of the proposed IoT system with goal-oriented communications and adaptive network resource management.}
    \label{fig:GOarchitecture}
\end{figure*}

In this section, we describe our envisioned GO IoT architecture. We consider multiple IoT devices, with possibly limited computational and energy capabilities, which are connected through one (or more) access points (APs) to a network of edge servers (ESs) endowed with a larger amount of computing resources; an illustration is given in Fig. \ref{fig:GOarchitecture}, where the three network segments interact with each other in a continuous and possibly seamless way. Generally, the higher we are in this hierarchical architecture, the more resources are available, but also higher delays are experienced to reach them. The objective of a GO system design and optimization is to strike the best trade-off in terms of resource accessibility, availability and cost, with the aim of fulfilling the desired goal. To perform an IoT task, the system handles several phases for each device, involving sensing, processing, communication, and control. The GO network management entity takes measurements from all network segments to monitor energy consumption and goal-effectiveness, and then solves GO optimization problems through the \textit{Goal-oriented optimization engine} (GO-OPT). A sketch of a typical problem solved by the GO-OPT is shown at the bottom left of Fig. \ref{fig:GOarchitecture}. The latter is driven by the key performance indicators (KPIs) reported at the top right of Fig. \ref{fig:GOarchitecture}. Then, a dynamic GO-OPT policy $\pi$ (which gives rise to a set of optimized connect-compute resources $\pi^*_t$ at time $t$) is fed back to all network segments (orange arrows) as inputs to allocate the needed resources. In the sequel, we describe the main elements that compose our GO IoT design, encompassing both device and network perspectives.

\subsection{Goal-oriented Communications at the IoT devices}

Referring to Fig. \ref{fig:GOarchitecture}, each IoT device is endowed with a \textit{goal-oriented communication module} (GO-COM) that is responsible to distill, extract, and transmit only the strictly relevant data features to pursue the communication goal. The GO-COM module encompasses the following functionalities:
\subsubsection{Effective sensing} Each device is equipped with (possibly multiple) sensors to acquire multi-dimensional signals, e.g., images, video, etc. \textit{Effective sensing} makes sure that only the relevant information is acquired \cite{kountouris2021semantics}. Indeed, when dealing with semantics and effectiveness, it is not only important how many (or how frequently) samples are collected, but also (and most importantly) which data samples are collected. The idea is to explore the spatial-temporal distributions of information, with the goal of detecting the effective relevance of different parts of data (across different dimensions) and targeting only the most important parts for sensing and acquisition. 

\subsubsection{Goal-oriented processing and communications} In this phase, there is an initial \textit{offloading decision},  which establishes whether the device prefers to perform the task fully locally, remotely (i.e., at the ESs) or partially at both ends by exchanging intermediate results. For instance, in the case of an inference task, the devices might decide that performing such operation locally (or partially) is more convenient than offloading it to the the network. 
Instead, whenever (part of) the task is offloaded, the GO-COM module extracts task-related features from data controlling the output dimension with a variable \textit{compression factor}. An \textit{encoding rule} then turns compressed features into a sequence of bits, even if fully analog designs are also possible. Depending on the task, the feature extraction and encoding can give different outputs. For inference purposes, initial GO designs proposed to adapt the size of the encoded data representation by acting on quantization or traditional compression schemes (e.g., JPEG) to obtain a target accuracy \cite{MerluzziEML2021,shlezinger2021deep}. More recent GO designs exploit deep learning architectures to extract compact features that are relevant for the communication goal \cite{xie2021deep,shao2021learning,jankowski2020wireless,binucci2023multi}.  For training purposes, the GO-COM module computes features related to a cooperative optimization task, e.g., stochastic gradients and/or local models, which can be further compressed to save communication resources. Hybrid solutions with partial offloading are also envisionable, e.g., through DNN splitting and/or early exiting. In any case, \textit {local CPU clock frequencies} must be allocated to perform the required computation, trading off latency, energy, and learning/inference performance. Finally, the devices transmit data features to the AP using \textit{GO adaptive modulation and coding}.

\vspace{-.2cm}
\subsection{Radio Access and Network Edge} 

In the radio access network, an \textit{assignment} procedure associates IoT devices with APs, selecting also the ESs that will take care of processing the data uploaded to the system. The assignment rule is dynamic, taking into account the mobility of users, the latency requirements of each application, and the availability of storage-connect-compute resources. A \textit{medium access control} protocol guarantees the co-existence between multiple users, considering also heterogeneous scenarios where goal-oriented and data-oriented communications coexist \cite{merluzzi20236g}. 
Once the selected AP receives data from its associated devices, it forwards the received messages to one (or more) ESs exploiting (wireless) high speed backhaul links. 
Each ES \textit{schedules its computing tasks} to fulfill the requests coming from different devices and the \textit{workloads are typically balanced} across the ESs (and, eventually, with the central cloud) depending on resource availability. 
Depending on the learning task, each ES either issues separate inference results for each device, or aggregates local estimates/models to perform ensemble or cooperative or learning (e.g., in the case of federated learning). 
In both cases, \textit {remote CPU clock frequencies} must be allocated per each served user by the ESs in order to perform the required computations. Finally, learning is typically exploited to implement an online control action, which can be performed either by the IoT devices (e.g., maneuvers for autonomous cars), or by the system itself for optimization purposes (e.g., network resource management). In both cases, the control action must be fed back to the devices through a dedicated control channel.

\section{Goal-oriented Adaptive Network Resource Optimization for IoT Services}


The first fundamental step needed to properly formulate GO communication problems is to formally define a goal. With the aim of embedding the GO perspective into IoT systems, we use the following definition \cite{merluzzi20236g}:\medskip\\
\fbox{\parbox{0.475\textwidth}{\textit{Definition:} A goal is the fulfilment of a task (e.g., learning, control,  actuation, etc.) characterized by a set of requirements that, if attained, determine its accomplishment.}} \smallskip\\
According to this definition, it is fundamental to define the main KPIs of the envisioned IoT system, which will determine the goal effectiveness requirements. In the sequel, we first introduce the basic concepts of goal effectiveness and goal cost; then, we will present an adaptive network resource optimization pursuing the system goal and encompassing several aspects such as sensing, computation, communication, learning, and control.

\subsection{Goal-oriented Key Performance Indicators}

In this section, we describe the main concepts of our IoT system design in terms of goal-effectiveness and cost.

\subsubsection{Goal Effectiveness} 
One of the fundamental aspects of goal-oriented communications is the definition of proper metrics to characterize the effectiveness level, i.e., a set of performance indicators that measure the accomplishment of an IoT task, e.g., image and/or video processing, vehicular/robotic control, etc. Each IoT application has its own performance metrics given by the task to be fulfilled. For instance, if inference services are required for classification or prediction purposes, the accuracy and/or the confidence of the inference task are key points to be considered for system optimization. 
However, as originally highlighted in \cite{MerluzziEML2021} and then refined in \cite{merluzzi20236g}, an inference/learning task is effective only if it is delivered within a given (possibly very stringent) delay specified by the application. For example, in vehicular IoT networks, it is of paramount importance to provide vehicles with both accurate and low-delay inference services (e.g., accurate image classification or cooperative perception within few milliseconds) to enable effective and safe control of the system. Similarly, for federated learning tasks, the GO effectiveness metric is typically related to the performance of the distributed training strategy within some specific latency constraints, e.g., convergence time and/or delay of single iteration \cite{battiloro2022lyapunov}. Thus, in very general terms, we define \textit{effectiveness as the simultaneous satisfaction of one or more task-related performance metrics coupled with E2E delay constraints}. A (non-exhaustive) list of goal-effectiveness metrics is given at the top right of Fig. \ref{fig:GOarchitecture}. We can define either short term (or instantaneous) and long-term effectiveness constraints considering, e.g., average, out-of-service, or probabilistic metrics.

\subsubsection{Goal Cost} Achieving a goal incurs a cost spent for its fulfillment, i.e., the price to be paid in terms of resources needed by fulfill the goal \cite{merluzzi20236g}. As an example, connect-compute energy consumption and spectrum utilization are typical costs to be taken into account in IoT system design. One of the key aspects of GO communications is the possibility to achieve the desired purpose of the communication while transmitting much less data with respect to conventional systems, thus achieving a substantial improvement in spectrum utilization. 
Our proposal is to use resources (e.g., energy and spectrum) only when and where needed 
to distil, communicate, and process only relevant information.  Clearly, in our envisioned IoT system, the overall cost is the sum of several components. From the devices point of view, the cost will include both local processing and uplink communication aspects. On the network side, the cost mainly comes from the activity of ESs and APs, related to edge processing and downlink communications (if needed). We can define either {\it short term} (or instantaneous) and {\it long-term} costs. In the latter case, we can define an average or a cumulative cost that consider all instantaneous costs spent to achieve the goal with the given effectiveness constraints.

\subsection{Goal-oriented Adaptive Resource Optimization}

In the envisioned scenario, there exist possibly several strategies achieving the goal with a target effectiveness, but generally leading to different costs. The aim of goal-oriented communications is to find the strategy that achieves the target goal-effectiveness with the lowest cost \cite{MerluzziEML2021,merluzzi20236g}. Both goal cost and goal effectiveness are affected by communication and computation aspects. For instance, GO compression schemes can be highly beneficial to reduce the cost and the E2E delay, while mildly trading goal effectiveness. Several interesting trade-offs between goal cost and effectiveness can be explored by adapting over time the available degrees of freedom depending on the dynamic system status. Referring to Fig. \ref{fig:GOarchitecture} (bottom left), GO communication problems can be naturally cast as a joint optimization of connect-compute resources, whose solution is an online policy $\pi_t$ ($t$ is the time-slot index) guaranteeing the target goal-effectiveness $\bar{\mathcal{E}}$ (a), subject to long-term (b) and short-term constraints (c), while paying the lowest possible goal cost $\mathcal{C}$. As previously mentioned, the GO-OPT is the architectural element that is responsible of carrying out the dynamic network resource management. This module takes on input goal-effectiveness measures from the IoT applications and physical world (e.g., accuracy, delay, reliability, etc.) and adapts over time the  network resource allocation $\pi_t$ to cope with the randomness of system parameters such as, e.g., wireless channels, data arrivals, availability of computing resources, and energy harvesting (if present). This is a new approach to network optimization, as the latter is usually performed based on predefined communication KPIs, and not on performance metrics reated to the application that motivated the exchange of data. The \textit{goal-oriented policy} $\pi_t$ is sent to all network elements and comprises sensing resources (e.g., sampling, source encoding), communication parameters (e.g., modulation and channel coding, transmit power, scheduling, user association), computation resources (e.g., computing resource scheduling at IoT devices and ESs), and learning modalities (e.g., ensemble, DNN splitting, federated operation), with the aim of achieving the target  effectiveness. Once the GO-COM of each device is updated with the new variables, new performance and effectiveness metrics will be measured and sent back to the GO-OPT, in order to adapt the resource allocation over time.

To tackle the complicated GO optimization problem, we resort to a mixed model-based and data-driven approach. We leverage Lyapunov stochastic optimization as a powerful method to learn optimal policies able to incorporate models describing the system operation, even in the presence of unknown statistics of channel state or requests arrival rates. This situation occurs in most of the communication and computation steps involved in several IoT tasks. Whenever a model is not available to quantify the impact of an action on a specific performance metric (e.g., how device selection affects FL accuracy), we exploit data-driven methods such as, e.g., deep reinforcement learning (DRL). 
Such interplay between model-based and data-driven optimization approaches 
endows IoT networks with powerful learning and adaptation capabilities, enabling the accomplishment of the goal with target effectiveness levels and minimum cost.


\section{Applications and Use-cases}

In the sequel, we show three use-cases of the proposed GO communications framework for IoT services, quantifying its performance in terms of trade-off between goal cost and effectiveness.

\subsection{Goal-oriented Compression for Edge Inference}

Let us consider a scenario where multiple devices aim to perform image classification by offloading the inference task to an ES through the wireless connection with an AP. Following the approach proposed in \cite{binucci2023multi}, the GO-COM module at the device is composed by a convolutional encoder (CE) that extracts the most relevant features for the classification task, while adapting the size of the encoded vectors to cope with dynamic variations of the wireless channel. Once offloaded, the compressed features are classified through a convolutional classifier at the ES. 
The GO-OPT minimizes the long-term energy consumption of the IoT system (i.e., the goal cost), while imposing long-term accuracy guarantees on the E2E delay needed to perform the task and the classification performance (i.e., the goal-effectiveness). We exploit Lyapunov optimization to devise an online resource allocation policy encompassing communication (e.g., compression, transmission rates, offloading decision) and computation resources (e.g., local and remote CPU frequency cycles, architecture selection). 

\begin{figure}[t]
    \centering 
    \vspace{-.4cm}
    \includegraphics[width=1.10\linewidth]{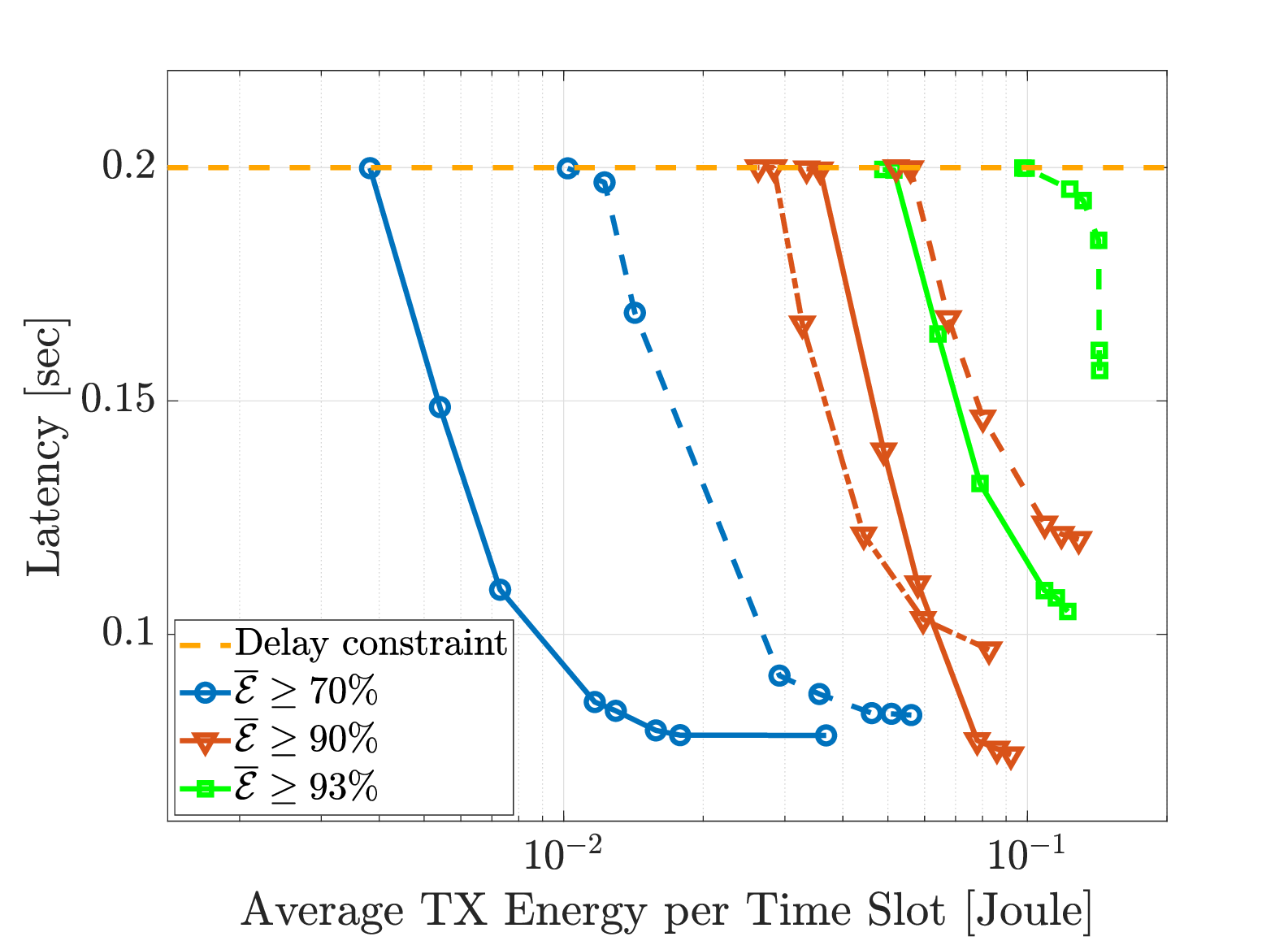}
    \caption{Energy/Latency trade-off. CE (solid) vs downsampling (dashed) for different minimum accuracy constraints.}
    \label{fig:binucci_energy_latency_trade_off}
\end{figure}

To give an example, we consider a scenario with 5 devices collecting 60 images/sec to be classified.
The maximum average E2E delay is set equal to 200 ms. Then, in Fig. \ref{fig:binucci_energy_latency_trade_off}, we illustrate the trade-off between (average) E2E delay and (average) system energy consumption, considering effectiveness targets (i.e., accuracy $\overline{\mathcal{E}})$. Our approach is depicted using solid curves and is compared with a more classical compression strategy, i.e., down-sampling with anti-aliasing pre-filtering, represented by the dashed lines. Furthermore, we also report the behavior of our GO optimization strategy when the GO-COM module is customized exploiting the GO data compression strategy in \cite{shao2021learning} (dashed dot line). From Fig. \ref{fig:binucci_energy_latency_trade_off}, as expected, the solutions with a lower energy consumption are also characterized by a higher latency and lower effectiveness. However, what is important to notice from Fig. \ref{fig:binucci_energy_latency_trade_off} is the large gain (around $2\times$) achieved by the proposed GO communication strategies in terms of cost-effectiveness trade-off with respect to conventional schemes, which is achieved thanks to the proposed GO system design.
\begin{figure*}[t]
     \centering
     \vspace{-.4cm}
     \begin{subfigure}[b]{0.49\textwidth}
     \hspace{-.2cm}
      \includegraphics[width=1.1\textwidth]{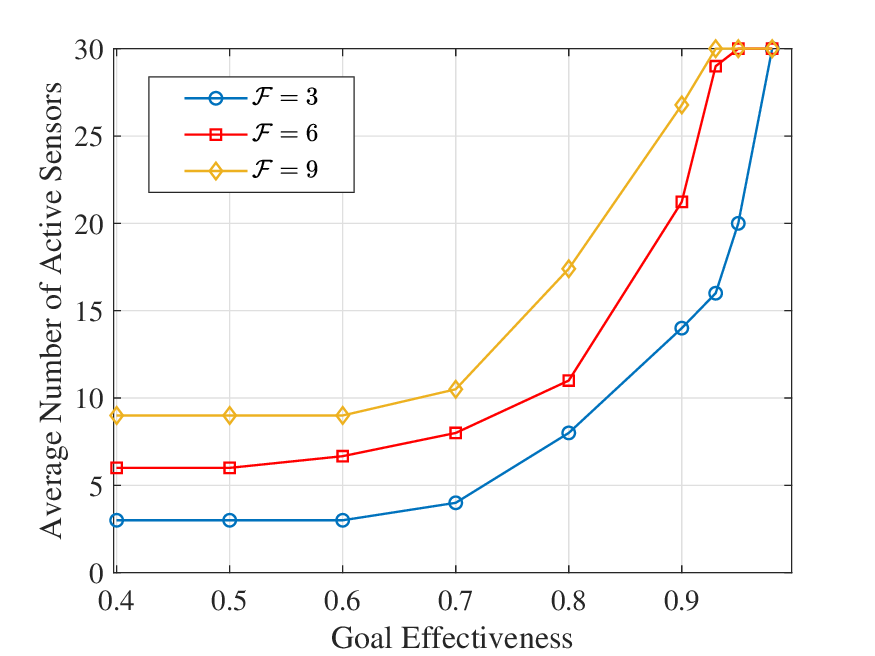}
         \caption{}
     \end{subfigure}
     \hfill
     \begin{subfigure}[b]{0.49\textwidth}
         \hspace{-.2cm}
         \includegraphics[width=1.1\textwidth]{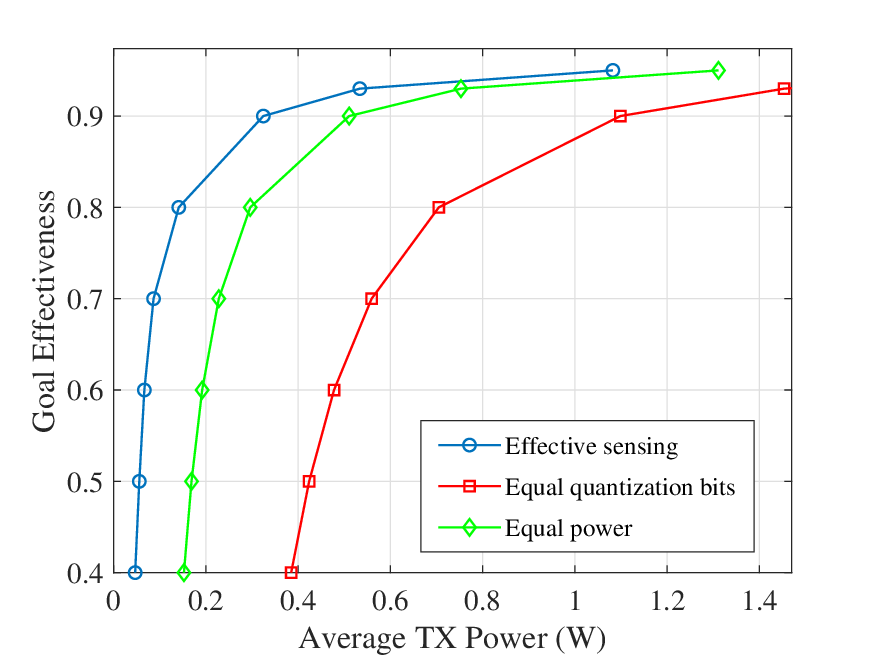}
         \caption{}
     \end{subfigure}
        \caption{(a) Average number of active devices versus goal effectiveness, for different subspace dimensions. (b) Goal effectiveness versus average transmission power, for different strategies.}
        \label{fig:cooperative_sensing}
\end{figure*}

\subsection{Cooperative Effective Sensing}

Let us consider a network of wireless devices gathering collected data towards a FC with the goal of evaluating optimal decentralized estimates of a signal of interest. 
If the data exhibits some kind of structure (e.g., correlations, smoothness, etc.),
which is typical in many physical fields of interest, a large compression can be achieved by performing GO data encoding in a joint and cooperative fashion. Differently from the case of Sec. IV.A, here we consider simple devices that can perform only low-complexity operations. Thus, the GO-COM module at the device is composed by an effective sampler, which collects data only if really needed, and by an adaptive quantizer that performs (joint) data compression. 
The GO-OPT module minimizes the instantaneous power consumption of the IoT system (i.e., the goal cost), while imposing accuracy guarantees on the estimation performance in terms of mean-square error (MSE) \cite{battiloro2020dynamic}. The goal effectiveness is measured as $\mathcal{E}=({\rm MSE}_{\max}-{\rm MSE})/{\rm MSE}_{\max}$, with ${\rm MSE}_{\max}$ denoting a maximum MSE value (set to $10^{-2}$) needed for efficient signal estimation. The output of the GO-OPT module is the number of quantization bits used by each device and the related transmit power. If the number of bits associated with one device is zero, the sampler does not collect the data.

To give an example, we consider a scenario with 30 devices, randomly placed over an area of 100 $m^2$, which observe a spatially correlated signal (with unitary norm) that is synthesized from the first $\mathcal{F}$ vectors of the Fourier basis. The observation noise is Gaussian i.i.d. over space and time, with zero mean and variance $10^{-4}$. The FC is placed at the center, and wireless channels consider path loss and Rayleigh fading with unitary variance. In Fig. \ref{fig:cooperative_sensing}(a), we report the average number of active devices (i.e., those collecting and transmitting data) versus the goal-effectiveness, considering 100 channel realizations, and 3 different values of the signal subspace dimension $\mathcal{F}$. As we can notice from Fig. \ref{fig:cooperative_sensing}(a), the number of active devices varies depending on the effectiveness constraint, increasing, as expected, when the requirements are more stringent. Interestingly, we can also observe that when the fields are more correlated (i.e., lower values of $\mathcal{F}$), the network requires a lower number of active devices to obtain a target effectiveness value. Also, for looser effectiveness values, the average number of active nodes tends exactly to $\mathcal{F}$, which is the minimum number of observations required for signal recovery. Finally, in Fig. \ref{fig:cooperative_sensing}(b), we report the  goal-effectiveness versus the average transmission power of the network, comparing the proposed effective sensing method with two baselines, which achieve the effectiveness constraint by letting all devices transmit with equal number of bits or power, respectively. As we can notice from Fig. \ref{fig:cooperative_sensing}(b), effective sensing leads to large gains in terms of cost-effectiveness trade-off with respect to baselines, showing how IoT networks can 
efficiently monitor a signal field of interest with the desired effectiveness.  





\begin{figure}[t]
    \vspace{-.4cm} \hspace{-.3cm}
\includegraphics[width=1.1\linewidth]{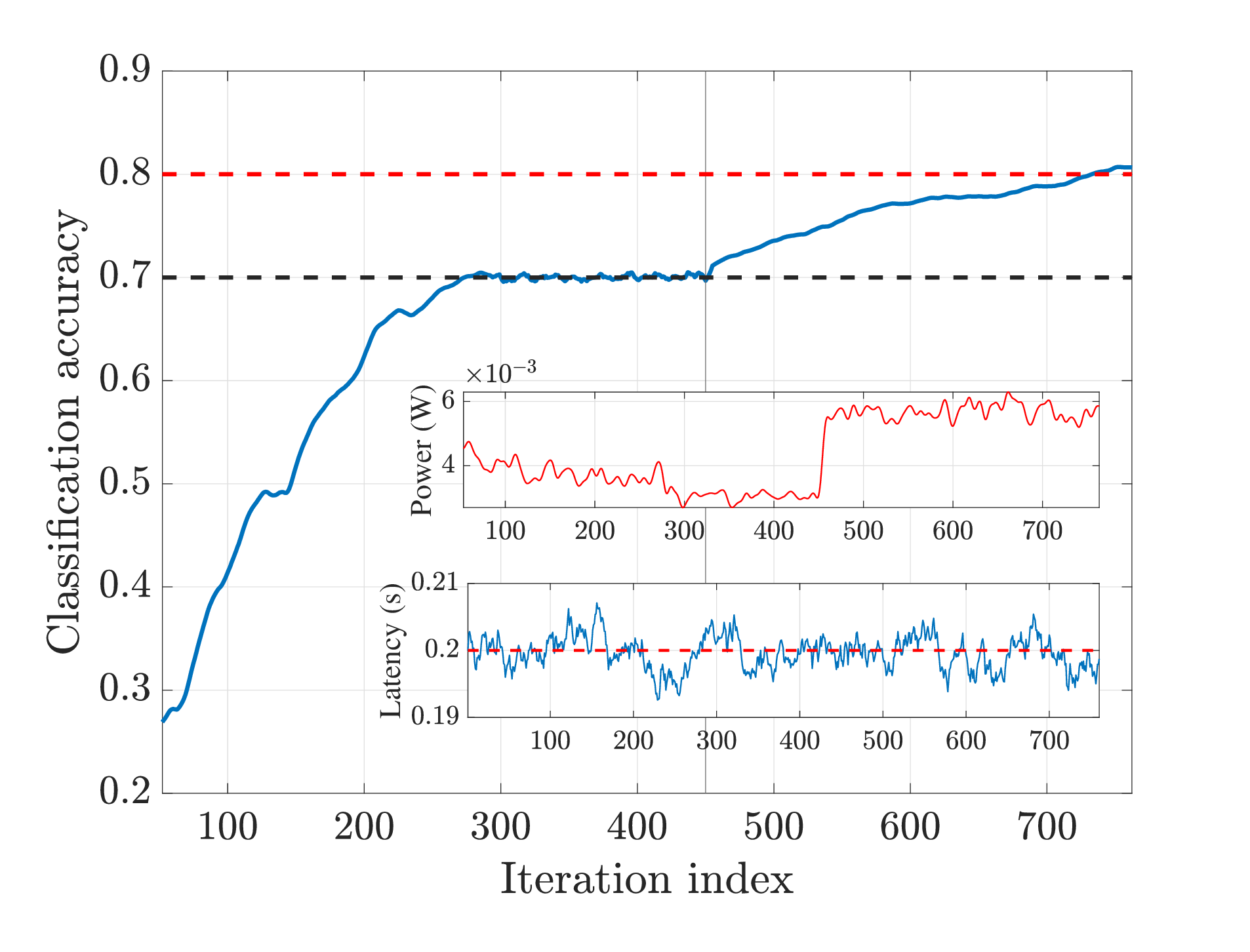}
\vspace{-.8cm}
\caption{Accuracy versus iteration index. Sub-plots: (Upper) Power versus iteration index, (Lower) Latency to perform one FL iteration over time.}
\label{fig:federated_learning}
\end{figure}

\subsection{Goal-oriented Federated  Learning}

Let us consider a set of IoT devices aims at training a common deep learning architecture with E2E delay and accuracy constraints, exchanging compressed models  computed from locally collected data. Following the approach proposed in \cite{battiloro2022lyapunov}, the GO-COM module at each device performs a back-propagation step from a batch of collected data to compute the local gradient. Such information is then used to produce a local model that is first quantized and then sent to the ES to produce a global model estimate by aggregating all the local models. Quantization level and batch size affect the FL performance, impacting on the quantity of noise injected into the local model computed by each device.
Also, the active set of transmitting devices plays a key role in discarding strugglers that would worsen learning performance because of bad available data or harsh wireless channels.  
The GO-OPT module minimizes the long-term power consumption of the IoT system, considering both communication and computation costs. The goal-effectiveness is expressed in terms of: i) long-term E2E delay constraints for the single FL iteration; ii) long-term constraints on the learning performance both in terms of accuracy (estimated online over an available set of data points) and convergence rate of the iterative FL algorithm. We exploit a Lyapunov-driven deep reinforcement learning framework to devise an online resource allocation policy encompassing communication parameters (e.g., quantization level, rates, set of active devices, etc.) and computation resources (e.g., CPU frequency cycles at the devices and the ES).

As an example, we exploit a convolutional neural network made of four layers, 
trained using the cross-entropy loss on the MNIST dataset.
During training, the GO-OPT module allocates resources to meet an accuracy constraint of 0.7 for the first 450 iterations, and of 0.8 afterwards. The constraint on the average latency per iteration is set to 0.2 seconds. In Fig. \ref{fig:federated_learning}, we report the instantaneous behavior of the classification accuracy on the test data, jointly with the corresponding power consumption and latency per iteration (upper and lower subplots, respectively). As we can see from Fig. \ref{fig:federated_learning}, the FL algorithm keeps learning until reaching the desired effectiveness level, while using the minimum power needed to guarantee the average latency constraint. At time 450, due to the new accuracy requirement, the algorithm is able to adapt to the new conditions, with the higher effectiveness paid in terms of a larger power consumption. This example shows the excellent adaptation and learning capabilities of the proposed GO design and adaptive optimization.

\section{Conclusions and Research Directions}


Goal-oriented communications represents a highly promising solution to address the scarcity of connect-compute resources, such as energy, spectrum, and computing power, and enable the envisioned IoT applications. This paper presents a goal-oriented system design for the IoT, focusing on both architectural and resource optimization aspects. By examining three case studies, we explore the benefits of our GO communication design and investigate the role of different degrees of freedom, such as sensing, communication, and computing resources, in achieving the trade-off between goal cost and effectiveness. Although preliminary, these insights are meaningful and have broad applicability in the IoT.

Several research directions are still open. As an example, it will be important to develop new information-theoretic fundamentals that encapsulate GO and/or semantic characteristics, together with representation of information in networked systems, to satisfy cost-related and effectiveness-related requirements in control, learning, and optimization. Another interesting aspect is the study of the interplay between cost, effectiveness, causality, and value of information in time-varying contexts, where the communication goal may change over time. Finally, GO and semantic principles should be exploited to devise innovative network resource management and control mechanisms, with the aim of improving sustainability and efficiency of future IoT networks.

\bibliographystyle{IEEEtran}
\bibliography{refs}



\end{document}